\documentclass[epsf]{aa}
\usepackage{natbib}
\usepackage{graphicx}

\def\cite#1{\citealt{#1}}

\def\NewA{New Astronomy}
\def\RMxAA{Rev. Mexicana Astron. Astrof.}
\def\ibvs{Inf. Bull. Variable Stars}

\begin{document}

\title{Recurrent Nova IM Normae}
\subtitle{}
\authorrunning{T. Kato et al.}
\titlerunning{Recurrent Nova IM Normae}

\author{Taichi Kato\inst{1}
        \and Hitoshi Yamaoka\inst{2}
        \and William Liller\inst{3}
        \and Berto Monard\inst{4}
}

\institute{
  Department of Astronomy, Kyoto University, Kyoto 606-8502, Japan
  \and Faculty of Science, Kyushu University, Fukuoka 810-8560, Japan
  \and Center for Nova Studies, Casilla 5022 Renaca, Vi\~{n}a del Mar, Chile
  \and Bronberg Observatory, PO Box 11426, Tiegerpoort 0056, South Africa
}

\offprints{Taichi Kato, \\ e-mail: tkato@kusastro.kyoto-u.ac.jp}

\date{Received / accepted }

\abstract{
   We detected the second historical outburst of the 1920 nova IM Nor.
Accurate astrometry of the outbursting object revealed the true
quiescent counterpart having a magnitude of $R$=17.0 mag and $B$=18.0 mag.
We show that the quiescent counterpart shows a noticeable variation.
From the comparison of light curves and spectroscopic signatures,
we propose that IM Nor and CI Aql comprise a new class of recurrent novae
bearing some characteristics similar to those of classical novae.
We interpret that the noticeable quiescent variation can be a result of
either high orbital inclination, which may be also responsible for the
low quiescent brightness, or the presence of high/low states.
If the second possibility is confirmed by future observations,
IM Nor becomes the first recurrent nova showing state changes in
quiescence.  Such state changes may provide a missing link between
recurrent novae and supersoft X-ray sources.
}

\maketitle

\section{Introduction}
   IM Nor was originally discovered as a possible nova in 1920 by I. E.
Woods from Harvard plates (cf. \cite{ell72imnor,due87novaatlas}).
The object was first detected on a plate taken on 1920 July 7 as
a 9 mag star.  Upon noting the possible
identification with a UHURU X-ray source 2U 1536$-$52\footnote{
The proposed identification with 2U 1536$-$52 = 4U 1538$-$52 was
already questioned by \citet{wyc79imnor}.  This X-ray source was later
identified with an X-ray binary, QV Nor (see \cite{kat02imnoriauc7791}).
}, \citet{ell72imnor} surveyed Harvard plates and constructed a light
curve.  Although the light curve was rather fragmentary, \citet{ell72imnor}
suggested a similarity with the light curves of slow novae, especially
DQ Her and the recurrent nova T Pyx.  No spectroscopic observation was
made during the 1920 eruption.

\begin{table*}
\begin{center}
\caption{Astrometry of IM Nor}\label{tab:astrometry}
\begin{tabular}{lccl}
\hline\hline
Source    & R.A. & Decl. & Remarks \\
          & \multicolumn{2}{c}{(J2000.0)} & \\
\hline
\citet{wyc79imnor} & 15 39 24\phantom{.465} & $-$52 19 34\phantom{.99} & 1,2 \\
\citet{due87novaatlas} (Harvard plate) 
                   & 15 39 26.25\phantom{5} & $-$52 19 21.3\phantom{9} & 1 \\
\citet{due87novaatlas} (candidate 1)
                   & 15 39 26.12\phantom{5} & $-$52 19 23.3\phantom{9} & 1 \\
\citet{due87novaatlas} (candidate 2)
                   & 15 39 26.50\phantom{5} & $-$52 19 22.2\phantom{9} & 1 \\
\citet{lil02imnoriauc7789}
                   & 15 39 26.61\phantom{5} & $-$52 19 18.6\phantom{9} & - \\
\citet{gar02imnoriauc7796}
                   & 15 39 26.465 & $-$52 19 17.99 & - \\
This work (from outburst image)
                   & 15 39 26.47\phantom{5} & $-$52 19 18.2\phantom{9} & - \\
Nearest USNO A2.0 entry
                   & 15 39 26.378 & $-$52 19 18.66 & 3 \\
DSS 2 quiescent counterpart
                   & 15 39 26.42\phantom{5} & $-$52 19 17.9\phantom{9} & 4 \\
\hline
\multicolumn{4}{l}{1: Precessed from B1950.0.} \\
\multicolumn{4}{l}{2: Remarked as incorrect in \citet{due87novaatlas}.} \\
\multicolumn{4}{l}{3: Blended.} \\
\multicolumn{4}{l}{4: Marked with tick marks in Fig. \ref{fig:id}.}
\end{tabular}
\end{center}
\end{table*}

   The quiescent identification of IM Nor was confusing.  At the position
of the nova, \citet{ell72imnor} remarked the presence of two stars
near mag 21.  \citet{wyc79imnor} spectroscopically studied these stars
and revealed that these stars are late-type stars without emission
lines.  \citet{wyc79imnor} concluded that the star is a late-type
companion of the nova, or that these stars are not physically associated
with the nova.  \citet{wyc79imnor} set an lower limit of $\Delta B$=11.7
as the outburst amplitude.  From this information and his original studies,
\citet{due87novaatlas} reported the remeasured position of the nova,
and different candidates for the quiescent counterpart, whose
exact identification remained uncertain (see Fig. \ref{fig:id}).

   The situation dramatically changed upon the discovery of the second
historical outburst by W. Liller (\cite{lil02imnoriauc7789}).
Subsequent observations confirmed the nova nature of the object
(\cite{due02imnoriauc7799}; \cite{ret02imnoriauc7818}).
This discovery qualifies IM Nor as the ninth recurrent nova in the Galaxy.

\section{Astrometry and Identification of the Prenova}\label{sec:prenova}
   Upon discovery of the outburst, we derived accurate astrometry from
a CCD image taken on 2002 January 15.104 UT with a 30-cm Schmidt-Cassegrain
reflector at the Bronberg Observatory near Pretoria, South Africa.
The astrometric reduction was done
using 66 GSC-ACT stars (mean residual $0^{\prime\prime}.3$).
Table \ref{tab:astrometry} gives
a summary of astrometry of this nova in the available literature.

\begin{figure*}
  \begin{center}
  \includegraphics[angle=0,width=7.5cm]{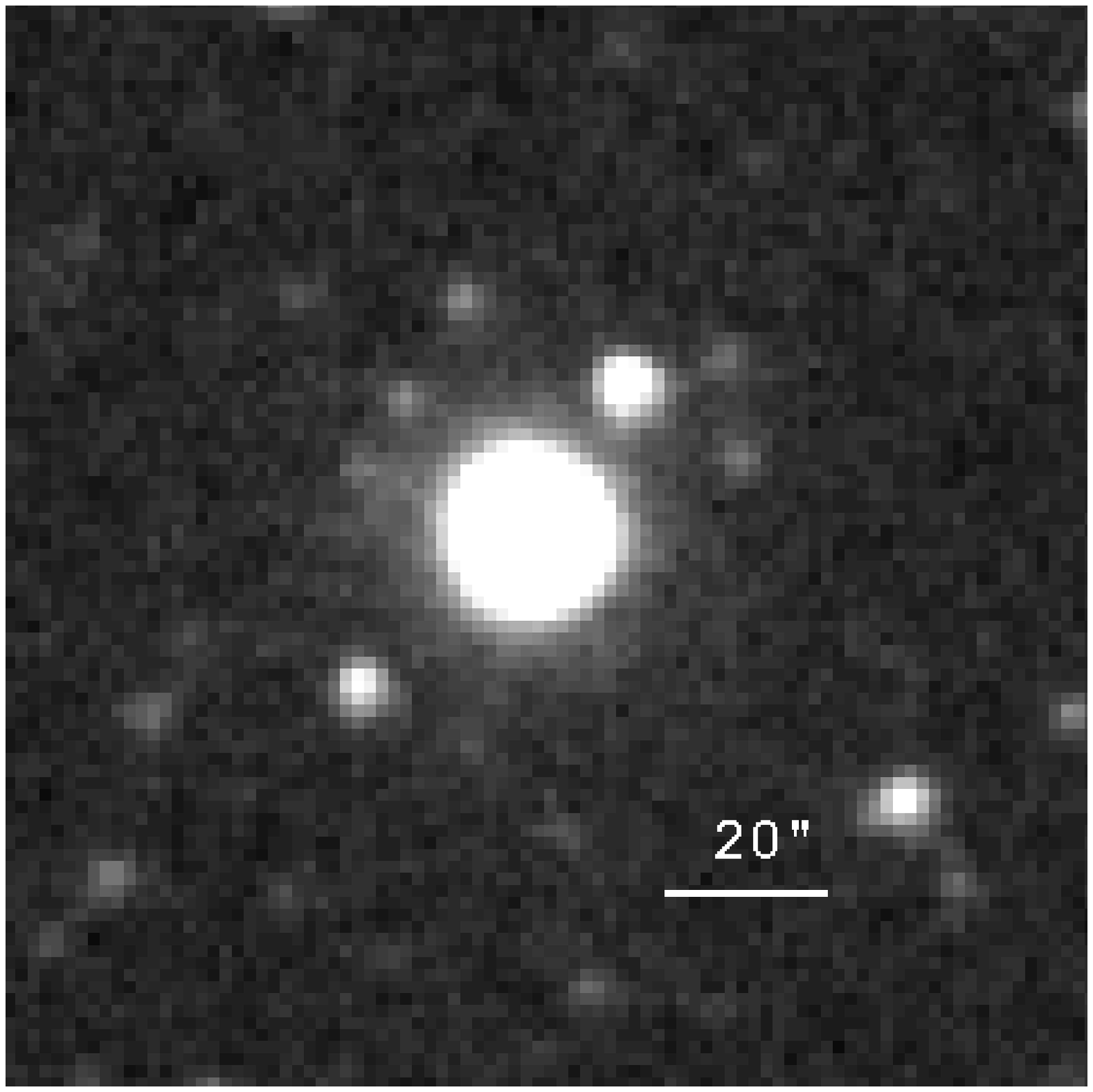}
  \includegraphics[angle=0,width=7.5cm]{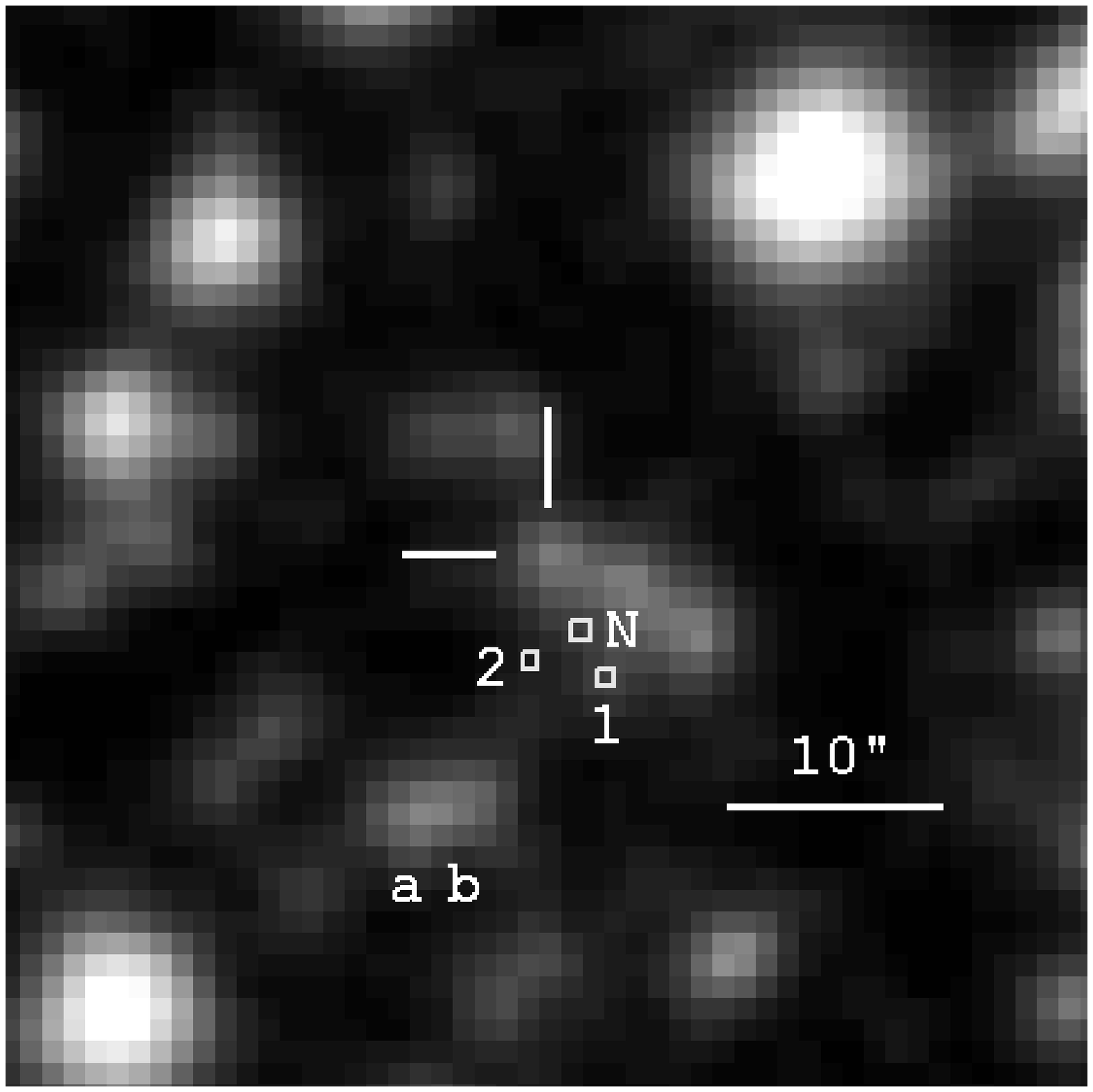}
  \end{center}
  \caption{Left: Outburst image of IM Nor.  Right: The quiescent
  counterpart of IM Nor (tick marks) on the DSS2 red image.  The stars
  labeled a and b are the proposed counterparts by \citet{ell72imnor}.
  The position of the nova and the proposed two candidate counterparts by
  \citet{due87novaatlas} are marked with squares labeled N, 1, 2, 
  respectingly (see Table \ref{tab:astrometry}).}
  \label{fig:id}
\end{figure*}

\begin{figure*}
  \begin{center}
  \includegraphics[angle=0,height=6.5cm]{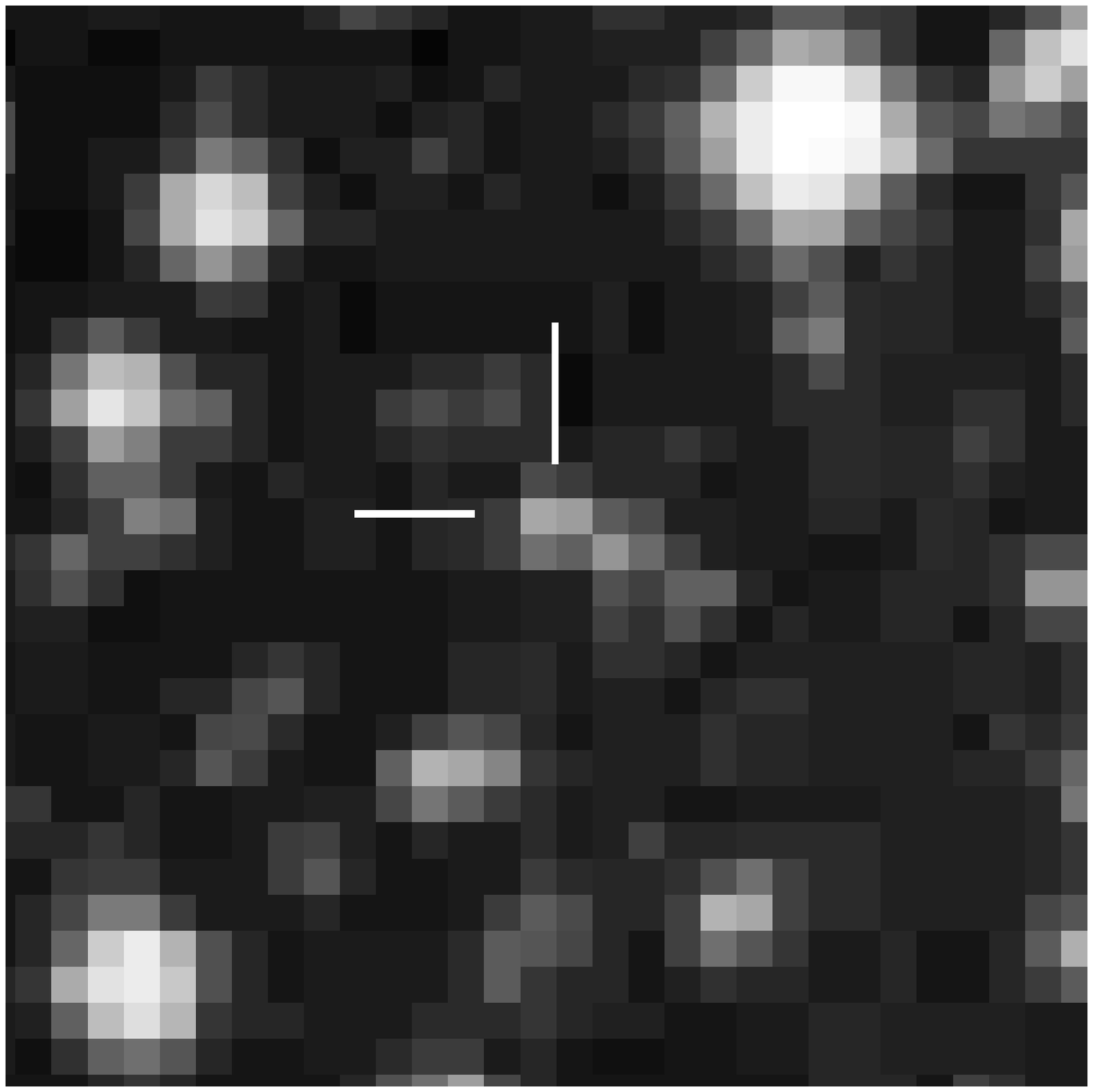}
  \includegraphics[angle=0,height=6.5cm]{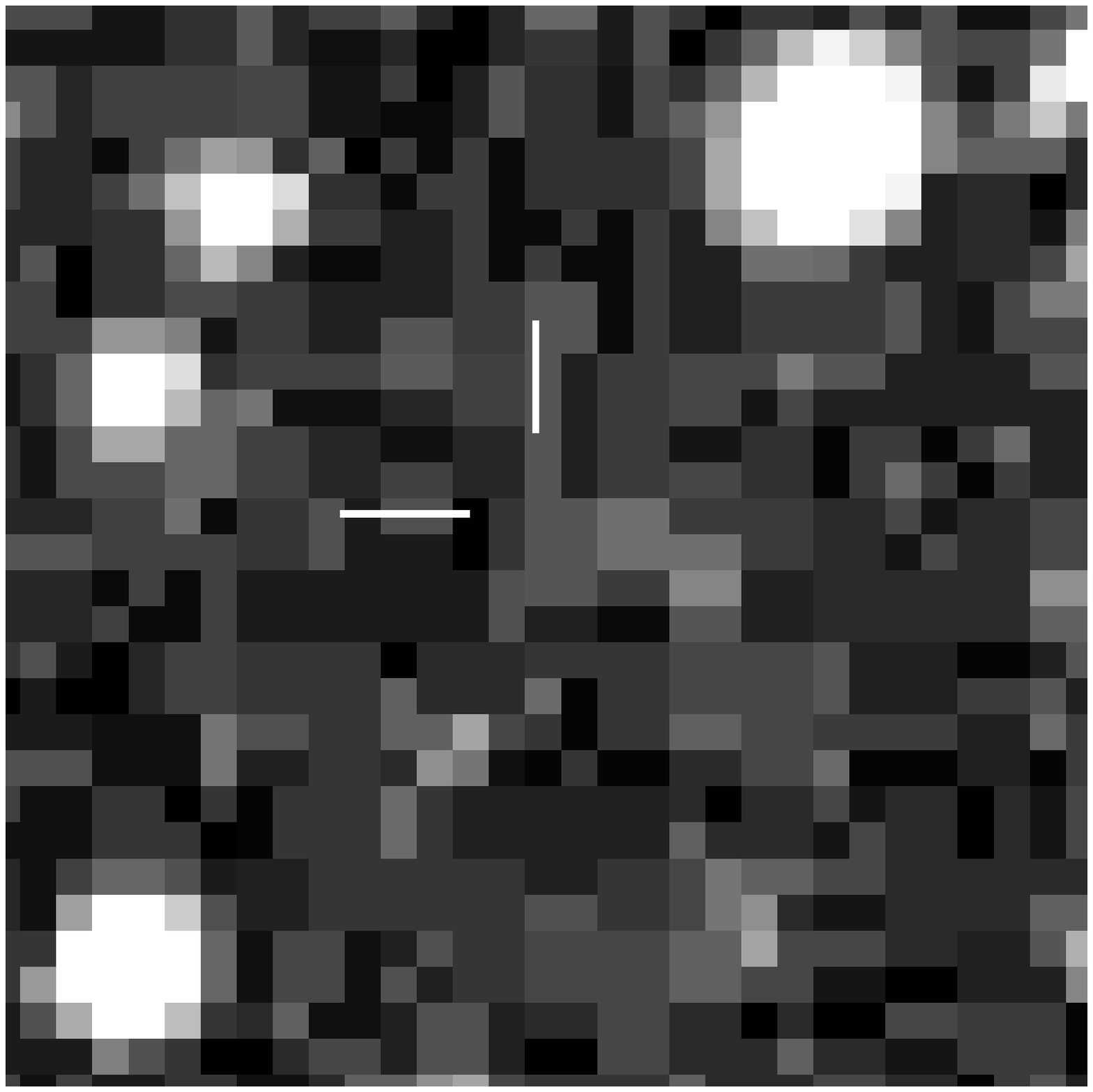}
  \end{center}
  \caption{Variation of the quiescent counterpart of IM Nor.  Left: The
  $B$-band image taken on 1975 Jul. 7 with UK Schmidt (the DSS1).  Right:
  The $V$-band image taken on 1987 Apr. 25 with the same telescope
  (also the DSS1).}
  \label{fig:quivar}
\end{figure*}

   Upon examination of the Digitized Sky Survey (DSS) images, we have
identified a star of $R$=17.0 mag and $B$=18.0 mag.
The star is different from the proposed
candidates by \citet{ell72imnor} (see also \cite{wyc79imnor} for a
finding chart) or by \citet{due87novaatlas}.  More detailed examination
of the available preoutburst photographs has revealed the presence of
noticeable variability (\cite{yam02imnoriauc7791}), which also strengthens
the prenova identification.  Estimated magnitude of the prenova are given
in Table \ref{tab:prenova}.  The correct identification, as well as past
suggested identifications, is shown in Fig. \ref{fig:id}.
This observation makes the observed amplitude of $\Delta B$ $\sim$10.0 mag.
A representative comparison of images is shown in Fig. \ref{fig:quivar}
demonstrating the presence of significant quiescent variation.\footnote{
  Even though the photometric bands are different between these observations,
  we can safely conclude that there was indeed a fading on 1987 April 25
  from the following reasons.  A $V$-band observation should naturally be
  brighter than $B$-band observation, considering the unavoidable
  interstellar reddening and the consistently observed positive $B-R$ color
  index.  The object is clearly visible in $R$ band image (Fig. \ref{fig:id})
  at the nearly same brightness as two western stars and also is in
  $B_{\rm j}$ image (Fig \ref{fig:quivar}a) brighter than these two.
  It is quite natural for the prenova.  The star, however, almost disappeared
  on the $V$ image (Fig. \ref{fig:quivar}b).  Such a change can
  not be explained by a color effect, but must have been caused by its
  intrinsic variability.
}

\section{Discussion}
\subsection{Classification among Recurrent Novae}

   From the relatively blue quiescent color, the object is unlikely a
symbiotic-type recurrent nova having a giant secondary
(T CrB, RS Oph, V745 Sco, V3890 Sgr: for recent reviews of recurrent novae,
see \cite{anu92RN,sek95RN,anu99RN,hac01RN}).
The object more resembles recurrent novae with main-sequence or
slightly evolved secondaries:
U Sco (\cite{bar79usco,sek88usco,sch90tpyxuscov394cra,sch95uscoperiod,
       hac00uscoburst,tho01uscomass});
V394 CrA (\cite{due88v394cra,sek89v394cra,sch90tpyxuscov394cra,hac00v394cra});
T Pyx (\cite{bar89cppuptpyx,sch90tpyxuscov394cra,sch92tpyx,kni00tpyx});
CI Aql (\cite{kis01ciaql,mat01ciaql,hac01ciaql});
Nova LMC 1999 No. 2 (\cite{sek90novalmc19902,sho91novalmc19902}).

\begin{table}
\begin{center}
\caption{Photometry of Quiescent IM Nor}\label{tab:prenova}
\begin{tabular}{cccc}
\hline\hline
Plate & Date & Magnitude & Band \\
\hline
J1628 & 1975 Jul.  7 & 18.0  & B  \\
XV224 & 1987 Apr. 25 & 19.5: & V  \\
XS224 & 1992 Jul. 24 & 17.0  & R
\end{tabular}
\end{center}
\end{table}

\begin{figure}
  \includegraphics[angle=0,width=8cm,height=10cm]{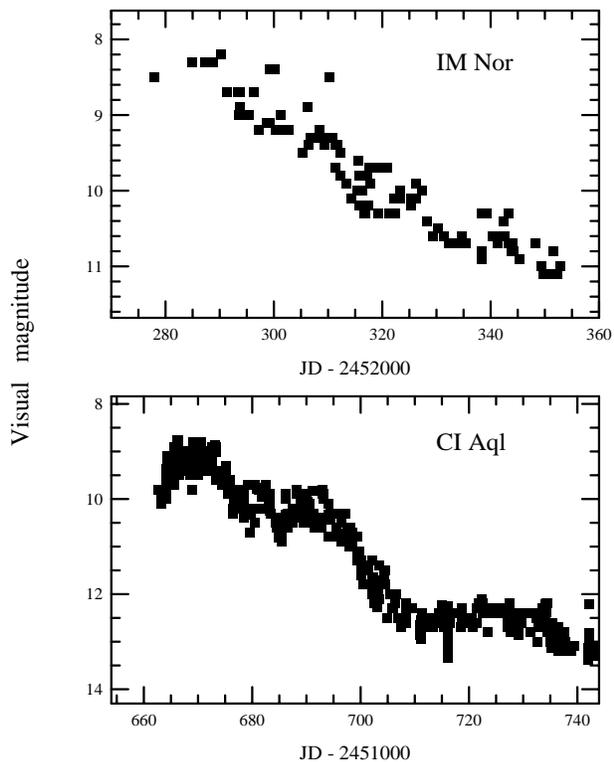}
  \caption{Comparison of light curves of the 2002 outburst of IM Nor
  and the 2000 outburst of CI Aql.  Modulations are superimposed on
  decays resembling those of moderately fast classical novae.}
  \label{fig:lc}
\end{figure}

Among these objects, U Sco (\cite{bar79usco}; \cite{sek88usco};
\cite{kiy99usco}), V394 CrA (\cite{due88v394cra}; \cite{sek89v394cra}
and Nova LMC 1990 No. 2 (\cite{sek90novalmc19902})
show extremely rapid decline, and are unlike IM Nor.
As already suggested by \citet{ell72imnor}, the slow decline of IM Nor
resembles that of T Pyx.  However, the light curve of the 2002 outburst
(Fig. \ref{fig:lc})
drawn from observations reported to VSNET Collaboration\footnote{
  http://www.kusastro.kyoto-u.ac.jp/vsnet/
} more suggests a moderately fast nova with a t$_3$ of $\sim$50 d.
The moderately structured light curve, in contrast to those of fast
recurrent novae (the best studied example being U Sco,
see \cite{hac00uscoburst}),
also makes a resemblance to a recently recognized recurrent nova,
CI Aql (\cite{kis01ciaql,mat01ciaql}).  From spectroscopy,
IM Nor is reported to show long persistence of Fe\textsc{II} lines
(\cite{ret02imnoriauc7818}), suggesting that more material has been
ejected than in a typical recurrent-nova outburst.  The spectroscopic
signature of massive ejecta makes a close resemblance to CI Aql which
showed a classical nova-like spectrum just after the maximum
(\cite{uem00ciaqliauc}) and the appearance of nebular lines at later stages
(Matsumoto et al., in preparation).  From these findings, we propose
that IM Nor and CI Aql comprise a new subclass of recurrent novae with
massive ejecta and long recurrence times.

\subsection{IM Nor in Quiescence}

    Quiescent IM Nor has several unique properties.  As shown in Section
\ref{sec:prenova}, the total outburst amplitude of IM Nor is
$\Delta B$ $\sim$10.0 mag, indicating that the quiescent counterpart is
fainter than usual recurrent novae (see e.g. \citet{hac00uscoqui}; the
large outburst amplitude of U Sco is difficult to reconcile without
a special mechanism).  Furthermore, IM Nor is unique among non-symbiotic
recurrent novae in that it apparently shows significant variation
in quiescence
(Section \ref{sec:prenova}; no high/low state transitions have been reported
in T Pyx, CI Aql and U Sco).  Such low quiescent brightness and
the presence of variations could be interpreted as an effect of
a high inclination, i.e. the observed variation in quiescence may
reflect eclipse-type or orbital variations.  In such a case, one may expect
to see orbital modulations or eclipses in near future (cf. U Sco:
\citet{hac00uscoburst}; CI Aql: \citet{mat01ciaql}).  We encourage
time-resolved photometry during the present outburst to test the
presence of possible eclipses, since eclipse observations during an outburst,
if observed, can severely constrain system parameters and the outburst
mechanism (\cite{hac00uscoburst,hac01ciaql}).  The low quiescent
brightness may also be a result of a circumbinary disk, as proposed in
U Sco (\cite{hac00uscoqui}).  Further observations of IM Nor in future
would provide a constraint to this interpretation.

   Alternately, if no eclipses are observed in IM Nor, the presence
of significant variation in quiescence makes the first indication of
high/low states in (non-symbiotic type) recurrent novae.
Since high/low state transitions
are more commonly seen in supersoft X-ray source
(RX~J0527.8$-$6954: \cite{gre96j0527SSSibvs,gre96j0527SSS};
CAL 83: \cite{alc97cal83,kah98cal83};
QR And: \cite{beu95qrand,gre95qrand};
V Sge: \cite{gre98vsgeSSS}), which are supposed
to be a close analog of recurrent novae
(\cite{kah99uscoSSS,gre00SSScatalog,hac99SNIaSSS}),
further observations and modeling of quiescent IM Nor is expected to
provide a missing link between recurrent novae and supersoft X-ray sources.

\vskip 3mm

The authors are grateful to the observers (Andrew Pearce, Jaime Garcia,
Raquel Yumi Shida, Bruce Tregaskis, Alexandre Amorim), who reported
visual observations of IM Nor to VSNET.
This work is partly supported by a grant-in aid (13640239) from the
Japanese Ministry of Education, Culture, Sports, Science and Technology.
This research has made use of the Digitized Sky Survey producted by STScI, 
the ESO Skycat tool, and the VizieR catalogue access tool.
The CCD operation of the Bronberg Observatory is partly sponsored by
the Center for Backyard Astrophysics.

\end{document}